\documentclass{ws-procs9x6}

\usepackage{epsfig}

\newcommand{\bmk}{\mathbf k}
\newcommand{\bmp}{\mathbf p}

\newcommand{\bmA}{\mathbf A}

\newcommand{\bmS}{\mathbf S}

\def\Dsl{\hbox{/\kern-.6000em D}} 

\def\dsl{\,\raise.15ex\hbox{/}\mkern-13.5mu D}
\def\bsigma{\mbox{\boldmath $\sigma$}}

\def\psip#1{\psi_{\mathbf{#1}}}
\def\chip#1{\chi_{\mathbf{#1}}}
\def\bsigma{\mbox{\boldmath $\sigma$}}

\def\abs#1{\left| #1 \right|}
\def\ltap{\ \raise.3ex\hbox{$<$\kern-.75em\lower1ex\hbox{$\sim$}}\ }
\def\gtap{\ \raise.3ex\hbox{$>$\kern-.75em\lower1ex\hbox{$\sim$}}\ }
\def\OMIT#1{}

\def\lsim{\mathrel{\raise.3ex\hbox{$<$\kern-.75em\lower1ex\hbox{$\sim$}}}}
\def\gsim{\mathrel{\raise.3ex\hbox{$>$\kern-.75em\lower1ex\hbox{$\sim$}}}}

\def\O#1#2{\mbox{\boldmath $O$}_{\mbox{\scriptsize\boldmath $#1$},#2}}

\begin{document}

\title{Summing Logs of the Velocity in NRQCD\\
 and Top Threshold Physics
\footnote{\uppercase{I}nvited talk at
 \uppercase{C}ontinuous \uppercase{A}dvances in  
  \uppercase{QCD} 2004, \uppercase{M}inneapolis, 
  \uppercase{USA M}ay 13-16, 2004.
\uppercase{P}reprint \uppercase{N}umber: \uppercase{MPP}-2004-153}
}  

\author{Andr\'e H.~Hoang}

\address{Max-Planck-Institut f\"ur Physik\\
(Werner-Heisenberg-Institut), \\
F\"ohringer Ring 6, \\
80805 M\"unchen, Germany\\
E-mail: ahoang@mppmu.mpg.de}

\maketitle

\abstracts{
To achieve reliable predictions of the top-antitop threshold cross
section at a future $e^+e^-$ Linear Collider logarithms of the top
velocity need to be resummed. I review the issues that make this
problem complicated and show how the task can be achieved
by renormalization in an effective theory using the so called
velocity renormalization group. The most recent NNLL order results are
discussed.} 

\section{Introduction}

The so-called ``threshold scan'' of the total cross section line-shape of top
pair production constitutes a major part of the top quark physics program at    
a future $e^+e^-$ Linear Collider.\cite{Teslanlc}  From
the location of the rise of the cross section a precise measurement of the top
quark mass with experimental uncertainties below 100~MeV will be
gained, while from the shape and the normalization of the 
cross section one can extract the top quark Yukawa coupling $y_t$ (for
a light Higgs), the top width or the strong coupling.\cite{TTbarsim}
 
In the threshold region, $\sqrt{s}\simeq 2m_t\pm 10$\,GeV, the top quarks
move with nonrelativistic velocity in the c.m.\,frame. Due to the large
top width, $\Gamma_t\approx 1.5$~GeV, toponium resonances cannot form
and the cross section is a smooth function of the c.m.\ energy. This
also means that non-perturbative effects can be neglected for
predictions of the total cross section.  
It is convenient to define the top velocity by $m_t
v^2\equiv\sqrt{s}-2m_t$. Because in the loop expansion one encounters
terms proportional to $(\alpha_s/v)^n$ from the instantaneous exchange
of $n$ time-like gluons, one has to count $v\sim\alpha_s$
and to carry out an expansion in $\alpha_s$ as well as in
$v$. Schematically the perturbative expansion of the R-ratio for the
total top-antitop threshold 
cross section, $R=\sigma_{t\bar t}/\sigma_{\mu^+\mu^-}$, has
the form
\begin{eqnarray}
 R & = & 
 v\,\sum\limits_{k=0}^\infty \left(\frac{\alpha_s}{v}\right)^k
\times
 \bigg\{\,1\,\,\mbox{(LO)}\,;
\, \alpha_s, v\,\,\mbox{(NLO)}\,;\, 
 \alpha_s^2, \alpha_s v, v^2\,\mbox{(NNLO)}\,\bigg\}
 \,.
 \label{RNNLOorders}
\end{eqnarray}
The expansion scheme in Eq.\,(\ref{RNNLOorders}) is called {\it fixed-order
expansion\,}, although it actually involves summations of the terms 
proportional to $(\alpha_s/v)^n$ to all orders. The scheme can be implemented 
systematically using the 
factorization properties of Non-relativistic QCD
(NRQCD) established by Lepage etal.~\cite{CaswellLepage}  
The NNLO QCD corrections 
to the total cross section were calculated already some time 
ago.\cite{Hoang1,Melnikov1,Yakovlev1,Beneke1,Nagano1,Hoang2,Penin1}
One of the important results was that in a threshold top quark mass scheme
(the most common schemes in use are the kinetic, PS and 1S mass
schemes\cite{BattagliaCKM}) the theoretical uncertainties for a top mass
measurement are comparable to the experimental uncertainties.
However, it was also found that  the NNLO corrections to the normalization of
the cross section were as sizeable as the NLO corrections leading to a
theoretical normalization uncertainty of at least $20\%$.\cite{Hoang3} 
It was shown~\cite{Peralta1,Hoang3} that this normalization uncertainty does
not affect significantly the top mass measurement since the latter is mainly
sensitive to the energy were the cross section rises. However, this
normalization uncertainty jeopardizes competitive measurements of the top
width, strong top coupling, and the top Yukawa coupling. 
Figure\,\ref{figtopplots}a  
shows the vector-current-induced R-ratio
$\sigma(e^+e^-\to\gamma^*\to t\bar t)/\sigma(e^+e^-\to\mu^+\mu^-)$ at LO, NLO
and NNLO in the 
fixed-order expansion for typical choices of parameters and 
renormalization scales.\footnote{
The contributions originating from Z-exchange are about an order of magnitude
smaller and can be neglected as far as discussions on QCD uncertainties are
concerned.} 
It is the discrepancy between the seemingly well
behaved NLO curves and the NNLO results which is the worrying issue.


One way to understand the large NNLO corrections is to recall that the
$t\bar t$ pair at threshold is non-relativistic and that its
dynamics is governed by vastly different energy scales, the top mass ($m_t\sim
175$\,GeV, ``hard''), the top three-momentum ($\bmp\simeq m v\simeq 25$\,GeV,
``soft'') and the top 
kinetic or potential energy ($E\simeq m v^2\simeq 3-4$\,GeV, ``ultrasoft''). 
The top width $\Gamma_t\approx 1.5$~GeV, which protects against
non-perturbative effects (and which is 
the main reason rendering the top threshold a precision
observable) is of order $E$, but can be neglected for most of the following
conceptual considerations. 
This hierarchy of scales is the basis of NRQCD
factorization,\,\cite{CaswellLepage} which separates short-distance
physics at the scale $m$ from long-distance physics at the 
non-relativistic scales $\bmp$ and $E$. In its original formulation NRQCD has
one quantum field 
for each of the non-relativistic quarks and antiquarks and one quantum field
for the low-energy gluons. 
NRQCD matrix elements and Wilson coefficients therefore involve three
types of logarithmic terms,
\begin{equation}
\ln\Big(\frac{\mu^2}{m_t^2}\Big)\,,\qquad
\ln\Big(\frac{\mu^2}{\bmp^2}\Big)\,,\qquad
\ln\Big(\frac{\mu^2}{E^2}\Big)\,.
\label{logarithms}
\end{equation}
These logarithms cannot be rendered small (or summed up) for any single choice  
of the renormalization scale $\mu$. However, this can be severe:
for example, $\alpha_s(m_t)\ln(m_t^2/E^2)\simeq 0.8$ for $\mu=m_t$ in the case 
of top quark pair production close to threshold. 
Obviously, a more sophisticated effective field theory (EFT)
framework has to be devised to resolve this issue and to allow to resum all
logarithmic terms. 


In this talk I discuss some of the conceptual aspects of an effective
theory that is capable of systematically summing all logarithms
of the velocity  that can appear in the description of the non-relativistic
quark-antiquark dynamics.  The tool needed is called the velocity
renormalization group (VRG)\cite{LMR} and it is the basis of a {\it
  renormalization group improved} perturbative expansion. To be specific, 
I will concentrate mostly on an EFT known as
vNRQCD.\,\cite{LMR,amis,amis2,Hoang4}. At the end I 
will come back to the top threshold cross section in order to review
the present status.

\section{The Proper Effective Theory}
\label{sectionvNRQCD}

To begin with, it is useful to summarize the properties one desires
from an effective theory of non-relativistic quark-antiquark pairs:
\begin{enumerate}
\item IR (on-shell) fluctuations, including IR divergences, in the
  full theory are reproduced by EFT fields. This avoids large
  logarithms in matching conditions.
\item There is a well-defined and systematic power counting scheme (in
  $v$). It can be uniquely identified which operators to account for what to
  compute at a given order. Moreover, all EFT loop integrations are governed
  by only one single scale. 
\item There is a consistent renormalization prescription to treat UV
  divergences and to formulate anomalous dimension which, eventually,
  will allow to sum all "large" logarithmic terms.
\item All symmetries (spin, gauge, etc.) are implemented. This gives
  the most predictive power.
\item The Lagrangian is formulated in the regulator-independent way. 
\end{enumerate}
The original NRQCD~\cite{CaswellLepage} does in fact not offer properties 
2 and 3. This is because there is one single gluon field describing soft and
ultrasoft gluon effects, while the power counting for soft and ultrasoft gluon
effects differs. For summing logarithms of the type shown above this is a
severe obstacle. For example, soft gluons are participating in the binding of 
the quark pair while ultrasoft gluons are responsible for retardation effects
such as the Lamb shift, and their interactions have to be multipole
expanded. The situation is, however, even more complicated
because, although $v\ll 1$ ensures that there is a strong hierarchy $m\gg
mv\gg mv^2$, the soft and 
ultrasoft scales turn out to be correlated by the energy-momentum relation of
the massive top quarks, $E\sim\bmp^2/m_t$. 

One approach in the literature to resolve these issues has been to ignore the 
correlation of soft and ultrasoft scales at the beginning and to account for
the scale hierarchy $m\gg mv\gg mv^2$ in the traditional Wilsonian (step-wise)
way. One starts 
with the NRQCD by Lepage etal. having soft gluons and then ``integrates out''
soft effects at the scale $mv$. This results in non-local heavy quark
interactions, the potentials, and interactions of the heavy quarks with the
remaining ultrasoft gluons. The theory below $mv$ is called
pNRQCD\cite{Pineda1} (p for ``potential'') and naturally avoids any double
counting of gluon effects. In the NRQCD-pNRQCD approach there is no unique
power counting. A v~counting taking into account the energy-momentum relation 
of the massive quarks exists in pNRQCD, while in the intermediate NRQCD theory a
$1/m$ counting is applied and the heavy quarks are static. A particular
property of pNRQCD is that the correlation of the soft and ultrasoft scales
has eventually to be implemented back into the theory when UV-divergences in
massive quark loops  occur. Renormalization group running caused by such
divergences then also couples the ultrasoft renormalization scale together
with the soft matching scale back to the hard scale $m_t$.\cite{Pineda2} 

In an alternative approach one accounts for the correlation of the soft and
the ultrasoft scales from the very beginning which then forbids using a
step-wise approach for the hierarchy of the momentum and the energy
scale. Rather, there is only one EFT below the scale $m_t$ which then contains
simultaneously soft and ultrasoft gluons. This approach is called
vNRQCD\cite{LMR,amis,amis2,Hoang4} (v for 
``velocity'') and allows for a 
consistent v~power counting that obeys the correlation $E\sim \bmp^2/m_t$ at
all scales below $m_t$. There is an ultrasoft 
renormalization scale, $\mu_U$, for loops dominated by the ultrasoft energy
scale, and a soft renormalization scale, $\mu_S$, for loops dominated by the
soft momentum scale which also includes the massive quark loops.
Both renormalization scales are  
correlated, $\mu_U=\mu_S^2/m=m \nu^2$. The running of coefficients and
operators in the EFT is then expressed in terms of the dimension-zero scaling
parameter $\nu$, where $\nu=1$ corresponds to the hard matching scale and
$\nu=v$ is the low-energy scale where matrix elements are computed. The
resulting renormalization group scaling, describing the  
correlated running coming from soft and ultrasoft effects is called the
VRG~\cite{LMR}. As long as one does not have to account for UV divergences 
coming from massive quark loops there is a rather straightforward
correspondence between the operator structure in vNRQCD and the one in the
NRQCD-pNRQCD approach.

The effective vNRQCD Lagrangian (defined in the c.m.\ frame) is build from  
heavy potential quarks and antiquarks ($\psi_\bmp$, $\chi_\bmp$), soft gluons,
ghosts, and massless quarks ($A_q^\mu$, $c_q$, $\varphi_q$) and ultrasoft
gluons, ghosts, and massless quarks ($A^\mu$, $c$, $\varphi_{us}$).
Double counting for the gluon effects is avoided since ultrasoft gluons
reproduce only the physical gluon poles where $k^0\sim 
{\bmk}\sim mv^2$, while soft gluons only have poles with $k^0\sim {\bmk}\sim 
mv$. All soft loops are made infrared-finite and at the same time all
ultrasoft divergences in ultrasoft loops are made correspond to the hard scale
$m$ by the pull-up mechanism.\,\cite{hms1}
It is essential that both soft and ultrasoft gluons are
included at all scales below $m$ because the heavy quark equation of motion
correlates the soft and ultrasoft scales. 
The dependences on soft energies and momenta of the heavy quark and soft gluon
fields appear as labels on the fields, while only the lowest-energy ultrasoft
fluctuations are associated by an explicit coordinate dependence. Formally
this is achieved by a phase redefinition
for the potential and soft fields\,\cite{LMR}, 
$ \phi(x) \to \sum_k e^{-ik\cdot x} \phi_k(x)$,
where $k$ denotes momenta $\sim mv$ and $\partial^\mu \phi_k(x)\sim mv^2
   \phi_k(x)$. 

\section{The Effective Theory Action}

The effective vNRQCD Lagrangian for a non-relativistic $t\bar t$ 
in its c.m.\,frame in a angular momentum S-wave and
color singlet state has terms\,\cite{LMR,amis,amis2}  
\begin{eqnarray}
\lefteqn{
\mathcal{L}   = 
 \sum_{\mathbf p}\bigg\{
   \psi_{\bmp}^\dagger   \bigg[ i D^0\!+\! i\frac{\Gamma_t}{2}
  \!+\!\delta m_t \!-\! {\left({\bf p}\!-\!i{\bf D}\right)^2
   \over 2 m_t} 
+ \ldots \bigg] \psi_{\bmp}
 + (\psi \to \chi)\bigg\}
}
\nonumber \\[2mm] && 
 -{1\over 4}G^{\mu\nu}G_{\mu \nu}
 - \mu_S^{2\epsilon}g_s^2 \sum_{{\bmp},{\bmp^\prime},q,q^\prime,\sigma} \bigg\{ 
  \frac{1}{2}\, \psi_{\bmp^\prime}^\dagger
 [A^\mu_{q^\prime},A^\nu_{q}] U_{\mu\nu}^{(\sigma)} \psi_{\bmp}
 + (\psi \to \chi) + \ldots\, \bigg\}
\nonumber
\end{eqnarray}
\begin{eqnarray}
&&
 + \sum_{p} \abs{p^\mu A^\nu_p -
 p^\nu A^\mu_p}^2 + \ldots
-\,\sum_{{\bmp},{\bmp^\prime}}
 \mu_S^{2\epsilon} V({\bmp,\bmp^\prime})\,\psi_{\bmp^\prime}^\dagger \psi_{\bmp}
   \chi_{-\bmp^\prime}^\dagger \chi_{-\bmp}
\nonumber \\[2mm] &&  
+\sum_{{\bmp},{\bmp^\prime}} \frac{2i\mu_S^{2\epsilon}\mathcal{V}_c}
  {(\bmp^\prime-\bmp)^4} f^{ABC}
  {(\bmp-\bmp^\prime)}.(\mu_U^\epsilon g_u {\bmA}^C) [\psi_{\bmp^\prime}^\dagger
  T^A \psi_{\bmp} \chi_{-\bmp^\prime}^\dagger \bar T^B \chi_{-\bmp} ] 
+\ldots
\,,
\label{vNRQCDLagrangian}
\end{eqnarray} 
where color and spin indices have been suppressed and $g_s\equiv g_s(m_t\nu)$,
$g_u\equiv g_s(m_t\nu^2)$, $g_s$ being the QCD coupling. All coefficients are
functions of the renormalization 
parameter $\nu$, and all explicit soft momentum labels are summed over. Massless
quarks and ghost terms are not displayed and there are other terms needed for
renormalization purposes\,\cite{Hoang5,Hoang6} not shown here. In
$d=4-2\epsilon$ dimensions powers of $\mu_U^\epsilon$ and $\mu_S^\epsilon$ are
uniquely determined by the mass dimension and the v counting of each
operator. The covariant 
derivative contains only the ultrasoft gluon field, 
$D^\mu = \partial^\mu+i\mu_U^\epsilon g_u A^\mu$. The first line in
Eq.\,(\ref{vNRQCDLagrangian}) contains the
heavy quark kinetic terms and their interaction with ultrasoft gluons.
There are
4-quark potential-like interactions of the form ($\bmk=(\bmp-\bmp^\prime)$) 
\begin{eqnarray}
 V({\bmp},{\bmp^\prime}) & = & 
 \frac{\mathcal{ V}_c}{\bmk^2}
 + \frac{\mathcal{ V}_k\pi^2}{m|{\bmk}|}
 + \frac{\mathcal{ V}_r({\bmp^2 + \bmp^{\prime 2}})}{2 m_t^2 \bmk^2}
 + \frac{\mathcal{ V}_2}{m_t^2}
 + \frac{\mathcal{ V}_s}{m_t^2}{\bmS^2}\,,
\label{vNRQCDpotential}
\end{eqnarray}
where $\bmS$ is the total $t\bar t$ spin operator. 

At NNLL order for the total
cross section the coefficient $\mathcal{ V}_c$ of the $1/\bmk^2$  potential has to be
matched at two loops\,\cite{hms1} because it contributes at the LL level, whereas
the coefficients of the order $1/m_t^2$ potentials only have to matched at the Born
level\,\cite{amis}. The $1/(m|{\bmk}|)$-type potentials are of order
$\alpha_s^2$ and have to be matched at two loops.\,\cite{Hoang5}
There are also 4-quark interactions  with the radiation of an
ultrasoft gluon  (last line) and interactions between quarks and soft gluons
(second line). Here, due to 
momentum conservation at least two soft gluons are required. The sum of the
potential terms shown in Eq.\,(\ref{vNRQCDpotential}) and time-ordered 
products of soft interactions contribute to the instantaneous interactions
between the top-antitop quark pair (that are traditionally called
``potentials'').

Particularly important for the phenomenological application are the effective
operators involving the top quark width and the residual mass term $\delta
m_t$ (first line).  The width term arises by including
the absorptive electroweak contribution from the W-b final state of the top
2-point function into the vNRQCD matching conditions at $\nu=1$. The width is
of order $\Gamma_t\sim E\sim D^0\sim\bmp^2/m_t\sim 
mv^2$ and has to be included in the heavy quark propagator. It is this width
term which explicitly serves as an IR cutoff and which causes the smooth cross
section shape visible in Figs.\,\ref{figtopplots}. Technically, it
corresponds to a shift of the c.m.\ 
energy into the positive complex plane by $i\Gamma_t$.\,\cite{Fadin1}
Beyond LL order, however, the structure of electroweak corrections is more
complicated.\cite{Hoang2,HoangReisser1} 
The residual mass term is non-zero in the threshold mass schemes, and cancels
the order $\Lambda_{\rm QCD}$ ambiguity arising from the higher
order QCD corrections to the Coulomb potential ${\mathcal{V}}_c/ \bmk^2$
that would otherwise destabilize the perturbative behavior of the location
where the threshold cross section rises.\cite{Hoang3} (In the pole scheme
$\delta m_t=0$, and the destabilization due to an 
${\mathcal{O}}(\Lambda_{\rm QCD})$ renormalon takes place.\cite{Hoang3})

Besides the interactions contained in the effective Lagrangian that describe
the dynamics of the $t\bar t$ pair we also need external currents that
describe the production of the top quarks. For $e^+e^-$ annihilation 
these currents are induced by the exchange of a virtual photon or a Z boson.
At NNLL order we need the vector S-wave currents  ${\bf
J}^v_{\bmp}= c_1(\nu) \O{p}{1} + c_2(\nu) \O{p}{2}$, where
$\O{p}{1} =  {\psip{p}}^\dagger\bsigma(i\sigma_2){\chip{-p}^*}$,
$\O{p}{2} = \frac{1}{m^2}{\psip{p}}^\dagger
    \bmp^2\bsigma (i\sigma_2){\chip{-p}^*}$ 
and the axial-vector P-wave current  ${\bf J}^a_{\bmp}=
c_3(\nu) \O{p}{3} $, where
$\O{p}{3} = \frac{-i}{2m}{\psip{p}}^\dagger
      [\bsigma,\bsigma\cdot\bmp](i\sigma_2)
   {\chip{-p}^*}$.
The currents $\O{p}{2}$ and 
$\O{p}{3}$ lead to contributions in the total cross section that are
$v^2$-suppressed with respect to those of the current $\O{p}{1}$.
Thus, at NNLL order, two-loop matching is needed for $c_1$ and Born level
matching for $c_2$ and $c_3$. 


\section{The Computations}

The computations for the total cross section in the EFT are performed in the
three steps, 
\begin{enumerate}
\item[i)] matching to QCD (including in principle also electroweak interactions) at
  $\mu_S=\mu_U=m$ ($\nu=1$);
\item[ii)] determination of the anomalous dimensions and running with the VRG from
  $\nu=1$ to $\nu=v_0$, were $v_0$ is the typical heavy quark velocity
  ($v_0\approx 0.15$ for $t\bar t$ threshold production);
\item[iii)] computation of the EFT matrix elements at $\nu=v_0$. 
\end{enumerate}
Conceptually, the novel aspect of the computations is the summation of
the logarithms of the correlated soft and ultrasoft scales using the
VRG. Although these calculations are usual standard once renormalization 
is expressed  
in terms of the renormalization parameter $\nu$, they are non-trivial
conceptually. This is because the  correlation of soft and ultrasoft
renormalization scales in the evolution is crucial to achieve the proper
summation of logarithms. This was demonstrated in QED,\,\cite{ManoharQED} where
higher order logarithmic terms could only be reproduced correctly using the
VRG. A nice demonstration at the technical level on the difference between
a correlated and an uncorrelated running was given in\cite{ManoharSoto}.
Technically the necessity of correlated running arises whenever divergences in 
massive quark loops {\it and} ultrasoft or soft loops 
contribute simultaneously to an anomalous dimension.
It was also shown\cite{Hoang4,Hoang6} that subdivergences in the NNLL (3-loop)
renormalization of the current $\O{p}{1}$, arising from diagrams where 
UV-divergences from massive quark and ultrasoft or soft loops arise
simultaneously in one single diagram, could be treated consistently in the
VRG, while an uncorrelated treatment of soft and ultrasoft renormalization
fails. 
Through steps i) and ii) all logarithmic terms involving $v$ are summed into
the operator coefficients and the matrix elements in iii) involve $\ln(v/\nu)$
terms that are not parametrically large.

As far as QCD effects are concerned, for the renormalization group improved
$t\bar t$ threshold cross section in vNRQCD all
necessary components in the steps i), ii) and iii) are known at NNLL
order with the exception of step ii) for the current coefficient $c_1$.
The running of the potential coefficients $\mathcal{V}_i$  
was determined in Refs.\,\cite{amis,hms1,amis3,Hoang5} and of the current
coefficients $c_i$ in Refs.\,\cite{amis3,Hoang5}.
Up to notational differences 
there is agreement with the potential coefficients obtained in
pNRQCD.\,\cite{PS1} 
The running of $c_1$ is fully known at NLL order,\,\cite{amis3,Hoang5} while at
NNLL order there is still some work to do.
The full set of non-mixing contributions coming from UV-divergences in actual
three-loop vertex diagrams have been determined in\cite{Hoang6}. A non-trivial cross
check for this computation, which checked the $\alpha_s^3\ln \nu$ term in
$c_1(\nu)$ was also carried out.\cite{Kniehl2} 
The full set of mixing
contributions, which arise from the next-order running of the Wilson
coefficients appearing in the NLL anomalous dimension of $c_1$ is still
unknown. Recently, the mixing contribution arising from the
spin-dependent ${\mathcal{O}}(1/m_t^2)$ potential (last term in
Eq.\,(\ref{vNRQCDpotential}) was computed in NRQCD-pNRQCD.\cite{Peninrecent}  


\section{The Top Pair Total Cross Section at Threshold}

\begin{figure}[b] 
\begin{center}
\leavevmode
\epsfxsize=3.8cm
\leavevmode
\epsffile[230 585 428 710]{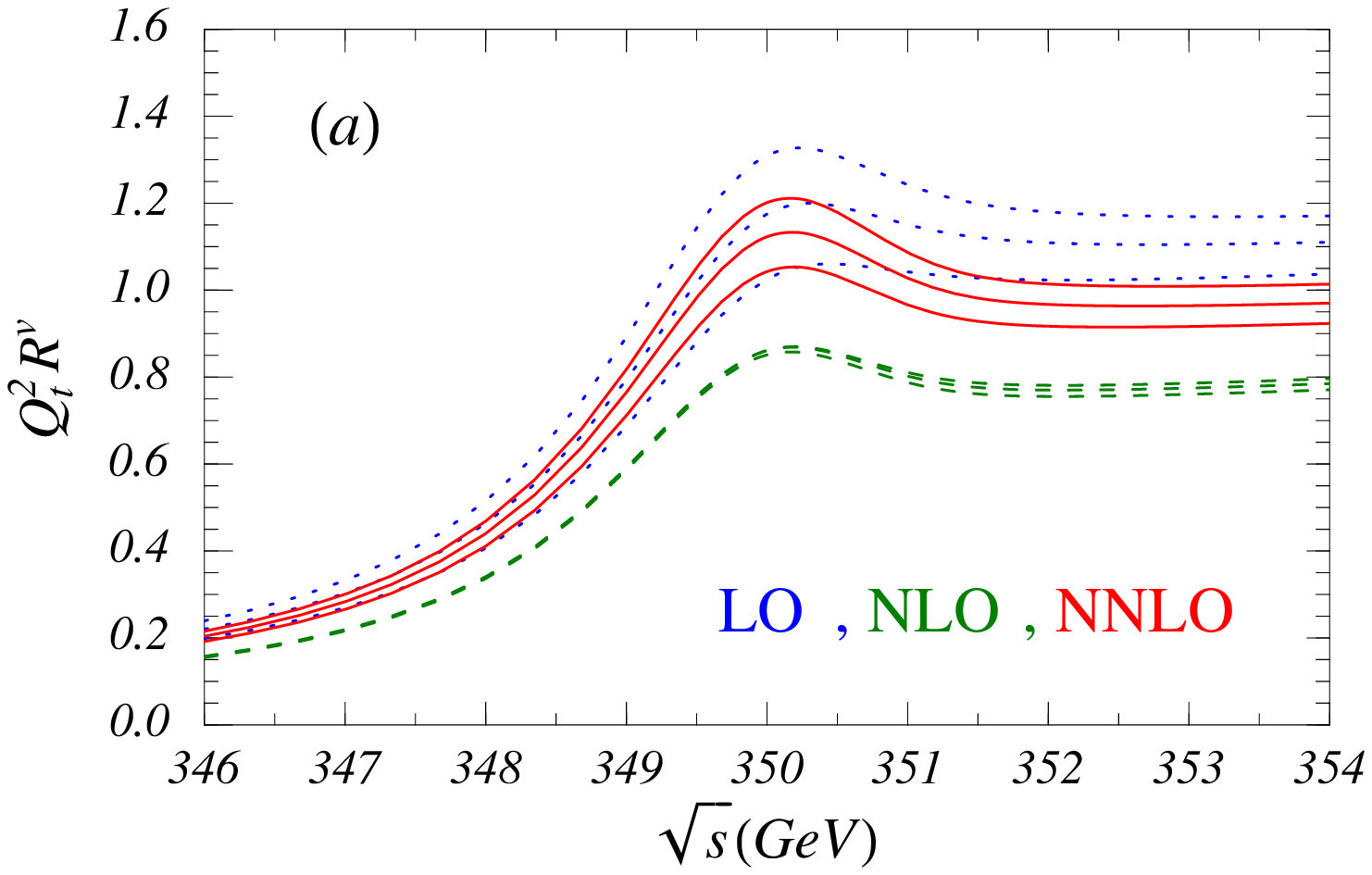}
\\[2.9cm]
\leavevmode
\epsfxsize=3.8cm
\leavevmode
\epsffile[230 585 428 710]{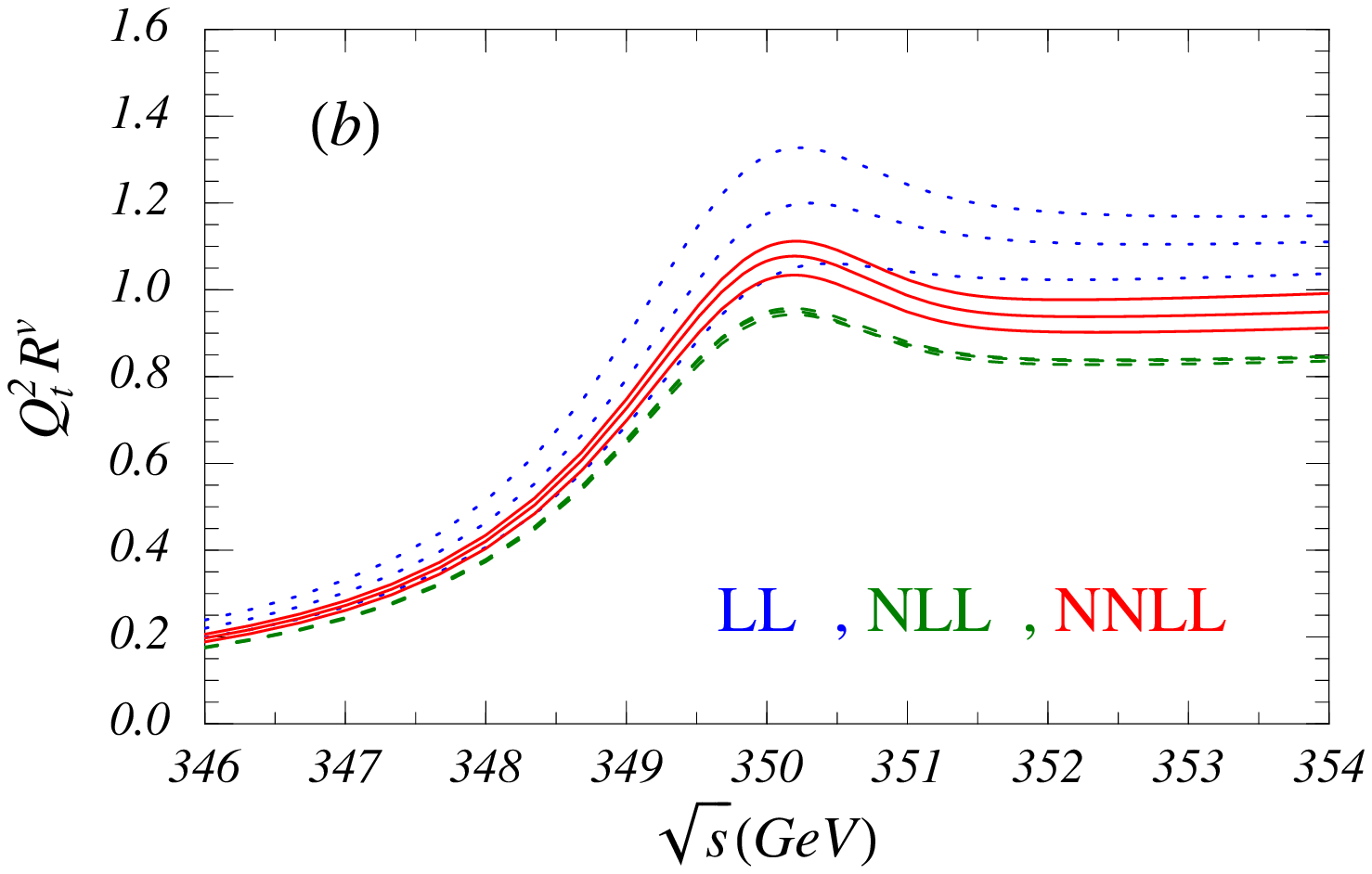}
\vskip  2.6cm
 \caption{ Panel a) shows the results for $Q_t^2 R^v$ with $M^{\rm
 1S}=175$\,GeV and $\Gamma_t=1.43$\,GeV in fixed-order perturbation theory at
 LO (dotted lines), NLO (dashed lines) and NNLO (solid lines). 
 Panel b) shows the results for $Q_t^2 R^v$ with the same parameters 
 in renormalization group improved perturbation theory at
 LL (dotted lines), NLL (dashed lines) and NNLL (solid lines) order. 
 For each order curves are plotted for $\nu=0.15$, $0.20$, and $0.3$. 
 The effects of initial state radiation, beamstrahlung and the beam energy
 spread at a $e^+e^-$ collider are not included. The plots are from Ref.[32]. 
\label{figtopplots} 
}
 \end{center}
\end{figure}
%
%

The total cross section for $e^+e^-\to \gamma^*, Z^*\to t\bar t$ at threshold
at NNLL order in renormalization group improved perturbation theory has the
form  
\begin{eqnarray}
  \sigma_{\rm tot}^{\gamma,Z}(s) = \frac{4\pi\alpha^2}{3 s} 
  \Big[\, F^v(s)\,R^v(s) +  F^a(s) R^a(s) \Big] \,,
\label{totalcross}
\end{eqnarray}
where $F^{v,a}$ are trivial functions of the electric charges and the isospin
of the electron and the top quark and of the weak mixing angle. 
The vector and axial-vector $R$-ratios have the form\,\cite{Hoang4} 
\begin{eqnarray} \label{Rveft}
 R^v(s) & = & \frac{4\pi}{s}\,
 \mbox{Im}\Big[\,
 c_1^2(\nu)\,\mathcal{ A}_1(v,m,\nu) + 
 2\,c_1(\nu)\,c_2(\nu)\,\mathcal{ A}_2(v,m_t,\nu) \,\Big] \,,
\\ \label{Raeft}
 R^a(s) & = &  \frac{4\pi}{s}\,
 \mbox{Im}\Big[\,c_3^2(\nu)\,\mathcal{ A}_3(v,m_t,\nu)\,\Big] \,,
\end{eqnarray}
where the $\mathcal{A}_i$'s are T-products of the effective theory
currents described before. As mentioned before, the running of $c_1$ is only
partly know at 
NNLL order. These T-products are related to the zero-distance
Green function of the equation of motion of the 4-quark 4-point functions
obtained in the EFT. (At NNLL order this is a common 2-body Schr\"odinger
equation, since higher Fock quark-antiquark-gluon states only contribute to
the renormalization of operators.) I use the 1S mass\,\cite{Hoang2,Hoang7}
since it is the threshold mass that is most closely related to the peak
position visible in the theoretical prediction. Other threshold masses are
also viable\,\cite{Hoang3} but not discussed here further.
  
In Fig.\,\ref{figtopplots}b I have displayed the photon induced cross section
$Q_t^2 R^v$ at LL (dotted blue lines), NLL (dashed green lines) and NNLL (solid
red lines) order for $\nu=0.15, 0.2$ and $0.3$.  Figure\,\ref{figtopplots}a
shows the corresponding results in fixed-order perturbation theory already
discussed earlier.
Compared to the fixed-order results with the same scales the improvement is
substantial, particularly around the peak position and for smaller
energies, but the somewhat inconvenient behavior that the NLL and the NNLL
order predictions do not overlap remains. At present the uncertainty of the
normalization of the total cross section is 
$\delta\sigma_{t\bar t}/\sigma_{t\bar t}\simeq \pm 6\,\%$ due
to this visibly large shift between the NLL and the NNLL order
predictions.\cite{HoangEpi} 
In fact, an uncertainty not larger than $3\%$ would be
required to have the theoretical errors of measurements of the top Yukawa
coupling, $\Gamma_t$ and $\alpha_s$ match with the ones presently expected
from experiments.\,\cite{TTbarsim}
It is, however, premature to draw definite conclusions as long as the NNLL
mixing contributions to $c_1$ have not yet been determined completely. 

\section{Acknowledgement}

I would like to thank the organizers of the conference 
for the hospitality and the pleasant atmosphere at the Theoretical Physics
Institute.

%
%
%
%


\end{document}